%% file: main.tex
\documentclass[%
 reprint,
superscriptaddress,
 amsmath,amssymb,
 prl,
 floatfix,
]{revtex4-2}

\usepackage{graphicx}
\usepackage{dcolumn}
\usepackage{bm}
\usepackage[english]{babel}
\usepackage[utf8]{inputenc}
\usepackage[squaren]{SIunits}
\usepackage[dvipsnames]{xcolor}
\usepackage{booktabs}
\usepackage{comment}
\usepackage{hyperref}
\hypersetup{
    colorlinks=true,
    linkcolor=blue,
    filecolor=magenta,      
    urlcolor=cyan,
}

\usepackage[normalem]{ulem}
\usepackage{cancel}

\newcommand{\jeti}{{JETi-200}~}
\newcommand{\ud}{\mathrm{d}}
\newcommand{\BCO}{BCO}

\begin{document}

\title{Polarization dependent beam pointing jitter in laser wake field accelerators}
\input{authors.tex}

\date{\today}

\begin{abstract}
We present experimental results, which show a laser polarization dependent contribution to electron beam pointing jitter in laser wakefield accelerators (LWFA). We develop a theoretical model for the polarization dependence in terms of the transverse dynamics of trapped electrons, resonantly driven by bubble centroid oscillations. The latter are generated by the carrier wave phase evolution
at the self-steepened laser pulse front. In the model, the polarization dependent jitter originates
from shot-to-shot fluctuations of the laser carrier envelope phase. The model is verified by particle in cell simulations and suggests that for non-CEP stabilized systems the polarization dependent jitter may form an ultimate limit to beam pointing stability in LWFAs.
\end{abstract}
                      
\maketitle

\paragraph{Introduction}

Laser wakefield accelerators (LWFA)~\cite{Dawson1979} have rapidly evolved from proof-of-principle to a stage where they combine extremely large accelerating fields with the generation of short electron bunches of few fs length \cite{lundh2011few,buck2011real,Debus2022CTR}, charge of  $\unit{100s}{\pico\coulomb}$ \cite{couperus2017demonstration} and low emittance~\cite{Barber2017_Emittance} with most recent work focusing on stability and reliability. A single cm-scale stage can reach energies of up to $\unit{8}{\giga\electronvolt}$ \cite{Gonzalves2019} with significantly higher energies possible in multi-stage systems. During the acceleration process a highly relativistic laser pulse is focused into a plasma, thus generating a plasma wave with strong longitudinal electrical fields of $\unit{100}{\giga\volt\per\metre}$ \cite{Esarey2009}. 
The transient nature of the plasma puts exacting requirements on the laser and plasma target to achieve the desired beam parameters. Controlling the phase-space properties of these transient, micron-scale accelerating structures is the focus of intense research. Major improvements have been achieved in  terms of the spectrum and the emittance of the electron bunch by developing schemes capable of controlling the injection process using schemes such as down-ramp-\cite{Chien2005_downram}, colliding pulse-\cite{faure2006colliding} and ionization injection\cite{Pak2010_ionization,McGuffey2010_ionization,Seidel2021_emittance}. These have achieved low energy spread, high energy, high charge and low emittance beams with high spectral stability, essential for a high quality accelerator. 
The stability of the transverse phase space, beam pointing and source position however, is equally critical. Experiments in strong-field QED\cite{Cole:PRX2018,Poder:PRX2018,SFQED-Review} or LWFA-based particle colliders  serve as examples,  with the most exacting experiments requiring a fluctuation much smaller than beam divergence and a variation of the source position of less than the source diameter. Up to now, such a degree of stability has not yet been achieved, and a physical understanding of the mechanisms deteriorating this transverse stability is still missing.

 It is well known that the electron bunch follows the laser pulse propagation to first approximation, since the wakefield and consequently the accelerating forces are  caused by the laser pulse. In all experiments to date, the pointing fluctuations have generally been attributed  to imperfections in the reproducibility of the gas target. The pointing fluctuations of modern laser systems can be as low as $\unit{2}{\micro \rad}$ \cite{Nakamura2017_laserPointing}. While substantially smaller than electron pointing fluctuations of $\unit{500}{\micro\rad}$ \cite{gonsalves2015generation}, they can become significant when amplifying effects due to density gradients are taken into account \cite{ma2018angular}.
Sporadic density ripples or shot-to-shot density fluctuations that occurred in early gas jet designs can contribute to additional jitter. However, in the capillaries and gas cells often used nowadays, these modulations are much smaller \cite{Kuschel2018} and can be neglected as a  source of jitter.
Other sources of pointing  jitter are pulse-front tilt \cite{Popp2010_PFT}, off axis injection of the electron bunch\cite{schnell2013optical,Lee_2015_betatron} and direct interaction of the laser with the electron bunch which is used to enhance hard X-ray generation\cite{huang2016resonantly_betatron} or fluctuations in the position of the laser near-field. These sources of pointing fluctuation can, in principle, be avoided under typical acceleration conditions. Importantly, they are not intrinsic to the acceleration process. The question arises whether there are any sources of pointing fluctuations which are an intrinsic instability of the acceleration process.
We note that in a homogeneous plasma, the laser generally causes a symmetric transverse density modulation due to the ponderomotive potential. For experiments with few-cycle pulses, however, where the rising intensity edge of the laser is very steep, an asymmetric transverse density modulation and further to pointing fluctuations due to carrier-envelope-phase (CEP) fluctuations of the laser have been observed in simulations \cite{Huijts:PhysPlasmas2021}. 
\begin{figure*}[t]
	\includegraphics[width=0.9\linewidth]{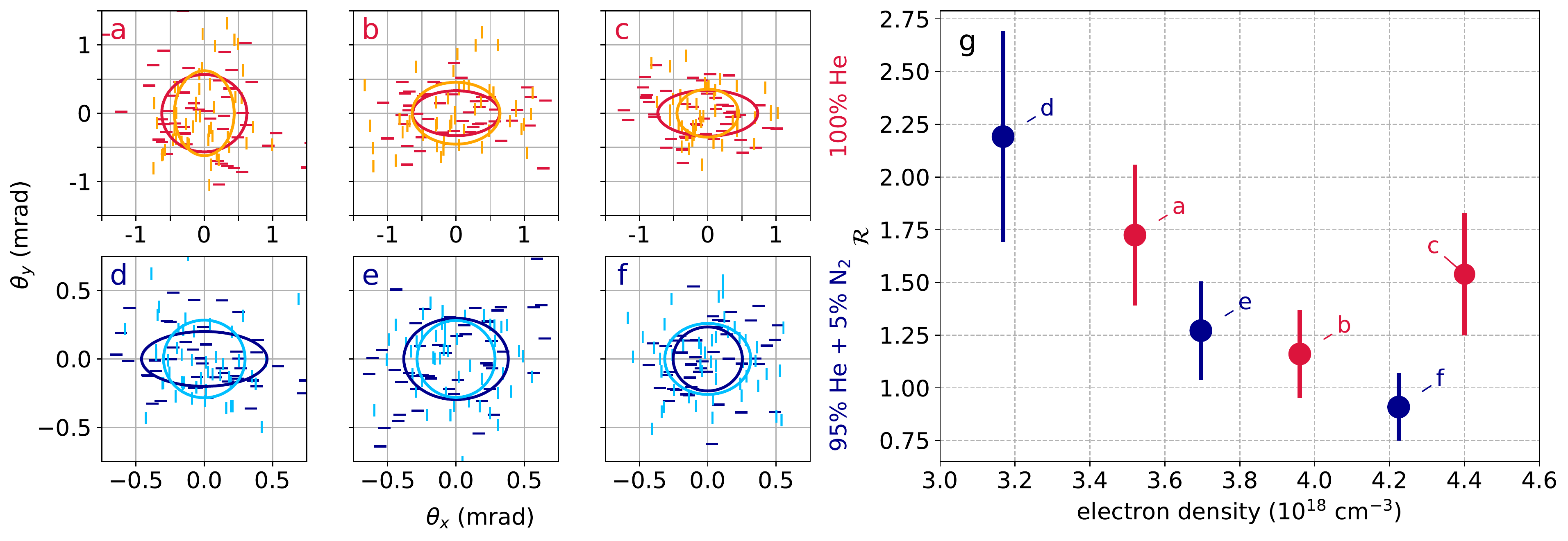}
	\caption{(a)-(f) Experimentally measured electron pointing data for vertical (vertical bars) and horizontal (horizontal bars) laser polarization. (a)-(c) pure helium (self injection), and (d)-(f) $95\%$ helium $5\%$ nitrogen mixture (ionization injection) at various plasma densities. Ellipses indicate the rms-jitter area of the electrons in respective polarization directions. (g) Magnitude of laser polarization induced electron pointing jitter, with $\mathcal R$ defined in Eq.~\eqref{eq:R}. All but  one data set show a positive correlation.}
	\label{fig:ExperimentalData}
\end{figure*}

Here we report on an instability affecting the ultimate limits of beam pointing jitter. We identify a mechanism {\it intrinsic} to LWFA coupling pointing in the polarization plane to CEP for the first time. This mechanism adds additional jitter in the direction of laser polarization and will be present even in the case of an ideal laser or target and is not restricted to few cycle laser pulses.  The polarization dependence of the beam jitter is explained by collective betatron oscillations resonantly excited by bubble centroid oscillations (\BCO{}s) due to the small intrinsic asymmetry of the ponderomotive expulsion of the background electrons.  For our conditions, the polarization induced jitter is of equal magnitude as the residual, polarization independent, contributions.

The experiment was performed at the \jeti laser facility at the Helmholtz Institute Jena, which  delivered in this experiment an energy up to $E=\unit{2.5}{\joule}$ on target at a center wavelength $\lambda_0 = \unit{800}{\nano\metre}$ and a pulse length $\tau=\unit{23}{\femto\second}$. Focusing with an f/24 off-axis parabolic mirror resulted in a vacuum FWHM focus diameter of $\unit{22}{\micro\metre}$ enclosing $\unit{40}{\%}$ of the total energy and consequently in a normalized vector potential of $a_0=2.5$.

The target was a length adjustable gas cell (kept fixed at $\unit{5.8}{\milli\meter}$ in this experiment) developed at Helmholtz Institute Jena. According to fluid dynamics simulations the gas flow from the entrance hole resulted in a $\unit{1.5}{\milli\meter}$ density up-ramp followed by a $\unit{5}{\milli\meter}$ constant density profile and a $\unit{1.5}{\milli\meter}$ down-ramp. The accelerated electrons were detected $\unit{187}{\centi\meter}$ downstream with a Kodak BioMAX scintillation screen \cite{kurz2018calibration}, allowing a precise measurement of the transverse electron beam charge distribution with a resolution of $\unit{0.05}{\milli\rad}$  by evaluating the center of a Gaussian fit to the charge distribution.
    
Figure ~\ref{fig:ExperimentalData} shows the experimental data for the electron pointing position of 50 consecutive shots taken with a linearly polarized laser interacting with pure He (a)-(c) and a $95\,\%$He $5\,\%$N$_2$ (d)-(f) gas mixture respectively. In each dataset the electron beam pointing jitter is shown both for horizontal and vertical laser polarization, which was controlled by a $\lambda/2$ waveplate. The polarization dependence is already visible in  the raw data and the ellipses depicting the rms-jitter of the electron pointing in horizontal and vertical direction, with the jitter generally larger in the direction of polarization.

In the following data analysis, it is assumed that the total electron pointing jitter $\sigma_{TX}$ can be attributed to a polarization independent part $\sigma_{0X}$ and a polarization dependent part $\sigma_P$ as following $\sigma_{TX}=\sqrt{\sigma_{0X}^2+\sigma_P^2}$, with $X$ indicating the horizontal ($H$)  or vertical ($V$) laser polarization direction. Experimentally, the $\sigma_{0X}$ and $\sigma_{TX}$ are accessible directly by rotating the input polarization. For horizontal polarization $\sigma_{TH}$ and $\sigma_{0V}$ are measured as the horizontal and vertical jitter component respectively. Conversely, for vertical polarization,  $\sigma_{TV}$ and $\sigma_{0H}$ are measured.
To avoid any time sensitive  drift, the data shots were collected consecutively for each plasma density in both horizontal and vertical laser polarization. The electron pointing data was evaluated using bootstrapping to calculate the mean jitter and uncertainty via the bootstrap distribution. Since $\sigma_{0X}$ depends on many variables, including plasma density and the specific electron injection mechanism,  only data with the same plasma density and injection mechanism were compared with each other. The experimentally measured components can be used to determine if the beam pointing exhibits a polarization dependence. The ratio 
\begin{align} \label{eq:R}
    \mathcal R = \frac{\sigma_{TH}}{\sigma_{0H}} \frac{\sigma_{TV}}{\sigma_{0V}}  = \frac{\sqrt{\sigma_{0H}^2+\sigma_P^2}}{\sigma_{0H}} \frac{\sqrt{\sigma_{0V}^2+\sigma_P^2}}{\sigma_{0V}} \,,
\end{align}
which is presented in Fig.~\ref{fig:ExperimentalData}(g) provides a measure for the relative polarization induced jitter.
In the absence of any polarization dependence the ratio $\mathcal R=1$, while $\mathcal R>1$ implies a larger  beam jitter in the laser polarization axis. Our data set shows a density dependent ratio for both self- and ionization-injection. Additional analysis using Bayesian Inference (see supplement \cite{supplement}) are in agreement with these findings.

Electron beam energy for the pure He gas reach a maximum  up to $\unit{550}{\mega\electronvolt}$ with a broad bandwidth of $\unit{300}{\mega\electronvolt}$ whereas electron spectra for the $95\,\%$He $5\,\%$N$_2$ gas mixture exhibit a peak around $\unit{600}{\mega\electronvolt}$ and a long tail extending below energies of $\unit{100}{\mega\electronvolt}$ (see supplement \cite{supplement}). At a plasma density of $n_e=\unit{4\cdot10^{18}}{\centi\metre^{-3}}$ the accelerating field is $\unit{150}{\giga\volt\per\metre}$ according to theoretical scalings \cite{lu2007generating,Esarey2009} implying an acceleration length of $\unit{4}{\milli\metre}$ in our experiment and consequently injection at the front of the cell for both injection mechanisms. 

Possible mechanisms that couple the laser polarization to the electron trajectories are (1) a polarization dependent injection mechanism \cite{Ma:PRL2020,Ma:PhysPlasmas2021,Kim:2021sub,Kim:2021yrb,Huijts:PhysPlasmas2021}, (2) electron interactions with the laser within the wake or (3) a coupling between the laser polarization and the bubble trajectory due to propagation effects in the plasma resulting the bubble oscillating as a whole. The first hypothesis was excluded by considering that the effect is present independent of injection mechanism as seen in  Fig.~\ref{fig:ExperimentalData} (a--c) (d--f). Both injection mechanisms follow a general density dependent trend (Fig.~\ref{fig:ExperimentalData}(g)),  although the absolute jitter $\sigma_T$ for self and ionization injection differ slightly. The second hypothesis can be rejected by noting that no signs of direct laser electron interaction in the electron spectra or the charge distribution were detected. Indeed the strongest polarization dependence is observed at low densities where the experiment is far from the dephasing length \cite{Esarey2009}. This trend is the opposite of what would be expected for the this hypothesis, where  the onset of the effect could only take place at a threshold density after which dephasing allows the electrons to interact directly with the laser.  We therefore exclude these two as a possible alternative explanation for the observed data.

The question therefore arises as to the origin of the observed polarization dependence of the beam pointing jitter. We recall that on a subcycle level, the electron motion is strongly dependent on the polarization: Each half cycle the electrons initially at rest at the front where the laser is creating the bubble by ejecting plasma electrons are primarily pushed to either one or the other side along the polarization axis. 
When only a few cycles are active in ejecting the electrons, the side receiving more electrons changes periodically due to the difference between phase and group velocity at the self-steepened laser pulse front as the laser propagates.
That this effect must exist can be seen by assuming a hypothetical pulse with a single cycle rising edge: the two ejection half-cycles must differ in strength. 
In the case of LWFA accelerators, self-steepening of the laser pulse front to a few cycle rise is the typical scenario of bubble formation. Significant front-steepness can be expected when the front-half of the pulse has been depleted, $L\sim \gamma_p^2 c\tau/2 \sim \unit{1.5 \ldots 2}{\milli\metre}$ \cite{Decker:POP1996,Schreiber:PRL2010}.

In the following we develop the theoretical model how the pointing jitter emerges from \BCO{}s due to self-steepened pulse fronts resonantly exciting collective electron betatron oscillations. Frequencies are normalized to the plasma frequency $\omega\to\omega/\omega_p$, distances are normalized to the plasma skin depth $x\to k_p x$, $\gamma_p^2 = n_\mathrm{crit}/n_e$; $\zeta=z-ct$.

    \paragraph{Bubble centroid oscillations.}

    Here we first investigate how \BCO{}s of the form $x_{bc}(t)=a_{bc}\sin(\omega_{bc} t+\phi_{bc})$ are excited. The direction of the ejection asymmetry is determined by the relative phase between the carrier wave and the steep pulse front. The laser phase changing by $2\pi$ at the pulse front equates to one period of the \BCO{}, and we can write $\omega_{bc} = \omega_L ( v_{ph} - v_{gr} - v_{etch})$ which yields an estimate for the oscillation frequency scaling as $\omega_{bc} \sim  \gamma_p^{-1}$ \cite{Nerush:PRL2009,Huijts:PhysPlasmas2021}.

    To estimate the amplitude $a_{bc}$ of the \BCO{} we have to consider the polarization-dependent corrections to ponderomotive momentum gain as first discussed by Ref.~\cite{Nerush:PRL2009} for the case of few-cycle pulses. The momentum gain due to the ponderomotive force $p^{(1)}(\pm x_0)\sim  \pm \frac{a_0^2}{\gamma_p w_0} \int \! f^2 \, \ud\zeta$, and the correction $p^{(2)}(\pm x_0)\sim \frac{a_0^3}{\gamma_p^2 w_0^2} \int \! f^2 g \, \ud \zeta$ have different symmetry for electrons initially at distance $\pm x_0$ from the beam axis. Here $f$ is the laser pulse profile, $a_x(x_0,\zeta) = a_0 f(x_0,\zeta)$, and $g$ its integral. While the ponderomotive force always pushes electrons away from the beam center, the correction $p^{(2)}$ deflects electrons into the same direction on both sides of the beam center, alternately each half cycle.

    As a consequence, the polarization dependent term $p^{(2)}$ slightly changes the average electron deflection angle by $\Theta \sim p^{(2)}/p^{(1)} \sim a_0 \hat F/\gamma_p w_0$, where the pulse front steepness parameter $\hat F = \max_{\phi_{CE}} F$ with $F = \int \! f^2 g \, \ud \zeta / \int \! f^2 \, \ud \zeta$. The function $F$ (cf.~supplement \cite{supplement}) is nonvanishing only if the temporal pulse profile $f$ changes rapidly on the wavelength scale. This is relevant, e.g., for few-cycle pulses \cite{Nerush:PRL2009,Huijts:PhysPlasmas2021}, or pulses which have undergone severe pulse-front steepening and thus acquired single-cycle features \cite{Ma:SciRep2016,Decker:POP1996,Streeter:PRL2018,Schreiber:PRL2010}. The \BCO{} amplitude can be estimated as $a_{bc} \sim \Theta R$, where $R=2\sqrt{a_0}$ is the bubble radius \cite{lu2007generating}, thus $a_{bc}\sim 2a_0^{3/2} \hat F / \gamma_p w_0$, and for a matched focal spot $w_0=R$ we find $a_{bc} \sim a_0 \hat F / \gamma_p$.

    \begin{figure}[!b]
        \centering
        \includegraphics[width=\columnwidth]{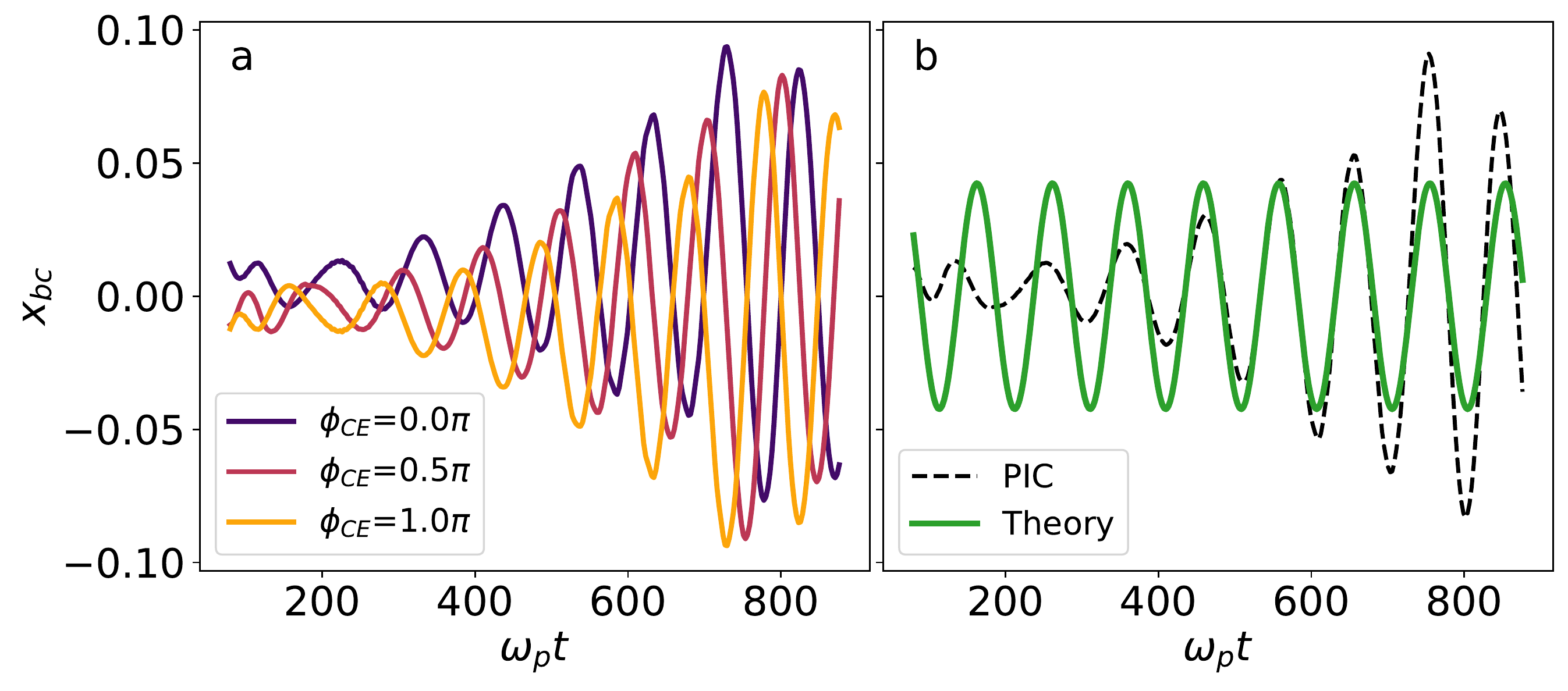}
        \caption{
        PIC simulation results showing the dependence of the \BCO{} on the laser CE phase (a) and comparison with the theoretical model (b) showing that the model correctly predicts the period, where $\omega_{bc}=0.064$, 
        $a_{bc} = 0.042 $ and $\phi_{bc}=1.2\pi$.  Note that the theoretical model assumes a static pulse-front steepness here while in the PIC simulation it evolves dynamically, resulting in the amplitude variation.
        }
       \label{fig:field_cep_dependence}
    \end{figure}

    In summary, the bubble centroid oscillates according to $x_{bc}(t) = a_{bc} \sin(\omega_{bc} t +\phi_{bc})$, where the scaling of amplitude $a_{bc}\sim a_0 \hat F / \gamma_p $ and frequency $\omega_{bc}\sim \gamma_p^{-1}$ follow from theory, but the phase $\phi_{bc} = \phi_{CE} + \phi_0$ is a-priori unknown since it contains contributions from the laser carrier-envelope-phase. To verify if the theoretically predicted oscillations are present in PIC simulations  the bubble centroid was defined in the latter via zero crossing of the transverse focusing field $E_x- B_y$ and the oscillation period of the model and simulations were seen to match ( Fig.~\ref{fig:field_cep_dependence}).

\paragraph{Betatron-\BCO{}-Resonance.}

    We now develop the model how \BCO{}s cause polarization dependent beam jitter via a (resonant) excitation of collective betatron oscillations of trapped electrons in the laser polarization plane. The latter are governed by the driven oscillator equation \cite{Mehrling:PRL2017}
    \begin{align} \label{eq:oscillator}
        \ddot x + \frac{\dot\gamma}{\gamma} \dot x + \omega_\beta^2 x = \omega_\beta^2 x_{bc}(t) \,,
    \end{align} 
    where both the Lorentz factor of the trapped electrons $\gamma$ and the betatron frequency $\omega_\beta(t)=1/\sqrt{2\gamma(t)}$ are slowly varying with time, with the \BCO{} $x_{bc}(t)$ as the driver. 
    As $\gamma$ increases with time, $\omega_\beta$ decreases. The maximum energy gain of a LWFA $\Delta \gamma \sim \frac{2}{3}a_0 \gamma_p^2$ \cite{lu2007generating} suggests that $\omega_\beta$ eventually drops below $\omega_{bc}$. Therefore, some time during the acceleration the betatron oscillations will become resonant with the \BCO{}, $\omega_\beta(t) \approx \omega_{bc}$. The \BCO{} varies the beam pointing angle $\theta=\dot x$ in the laser polarization direction, and the pointing jitter is caused by \BCO{} phase variations due to shot-to-shot CEP fluctuations. This we now demonstrate within our model.

    The WKB solution of Eq.~\eqref{eq:oscillator} is given by
    \begin{align}    \label{eq:x_ebeam}
    	&x (t) = \sqrt{\frac{\omega_{\beta}(t)}{\omega_{\beta}(t_i)}} \left(x_0 \cos[\varphi(t)]  + \frac{ \theta_0 }{\omega_{\beta}(t_i)}  \sin [\varphi(t)] \right)  \\
    	&+ \sqrt{\omega_{\beta}(t)} \int_{t_i}^t \sqrt{\omega_{\beta}(t')} 
    	        \sin \left[ \varphi(t)-\varphi(t^{\prime}) \right ] 
    	        x_{bc}(\phi_{bc},t^{\prime})) \: dt^{\prime} \nonumber 
    \end{align}
    with $\varphi(t) = \int_{t_i}^t \: \omega_{\beta}(t^{\prime}) \, dt^{\prime}$, where the second line is collective betatron oscillation driven by the \BCO.

    If we assume the injection time interval is short compared to the betatron and \BCO{} periods, it is sufficient to consider the beam centroid trajectory in order to get a qualitative understanding of the pointing jitter due to the betatron-\BCO-resonance. For a centrally injected beam ($\bar{x}_0=\bar\theta_0=0$), Eq.~\eqref{eq:x_ebeam} shows that the \BCO{} excites collective betatron oscillations with instantaneous pointing angle
    \begin{align} \label{eq:angle}
    \theta = \omega^{3/2}_\beta(t) \int_{t_i}^t \sqrt{\omega_\beta(t')} \cos \left[ \varphi(t)-\varphi(t^{\prime}) \right ] x_{bc}(t^{\prime}) dt^{\prime} \,.
    \end{align}
    To calculate the pointing jitter we now average over the unknown fluctuating phase $\phi_{bc}$ (denoted by $\langle\ldots\rangle$) and immediately find that the 2nd line of Eq.~\eqref{eq:x_ebeam} vanishes. Thus, irrespective of the initial conditions of the injected beam's centroid, $\theta - \langle\theta\rangle$ is given by Eq.~\eqref{eq:angle}.

    \begin{figure}[!t]
    \centering
    \includegraphics[width=0.85\columnwidth]{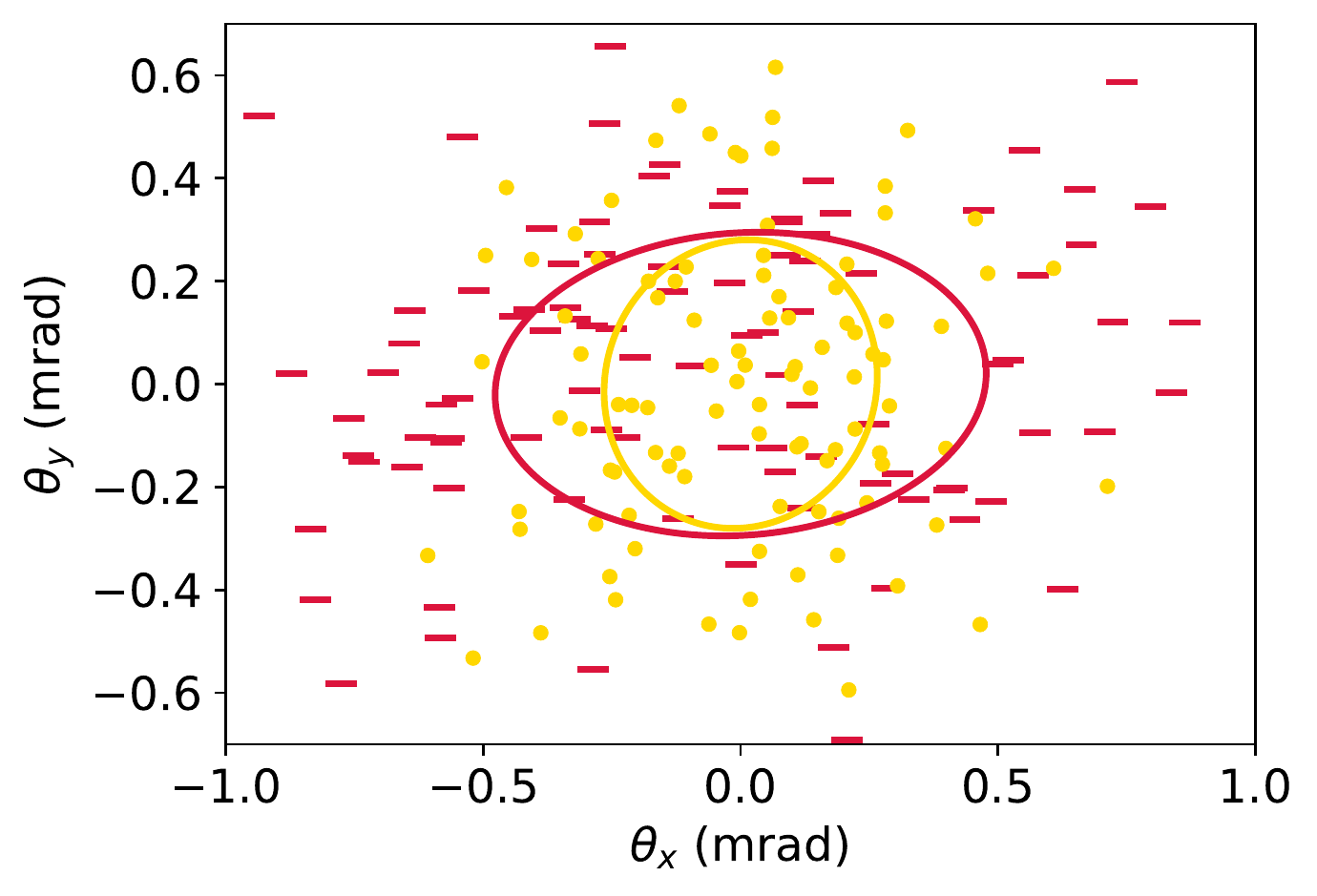}
    \caption{The pointing jitter calculated using the betatron-bubble resonance model is in agreement with the experimental data shown in Fig.~\ref{fig:ExperimentalData}. Yellow symbols are without \BCO{}s, i.e.~pure betatron oscillation with $x_{bc}=0$ and no polarization effect, while red symbols are for a horizontally polarized \BCO{}s with $a_{bc}=0.005$, $\omega_{bc}=0.03$, and the phase $\phi_{bc}$ uniformly distributed. The initial position of the beam centroid was normally distributed to account for the polarization independent jitter.}
    \label{fig:synthetic}
    \end{figure}

    The pointing fluctuation is determined by the second moment $\Delta \theta^2 := \langle (\theta-\langle\theta\rangle)^2\rangle$,
    \begin{align} \label{eq:jitter}
    \Delta \theta^2 = \frac{\omega_\beta^3(t)}{8} 
    \left| \int_{t_i}^{t} \! dt' \: \sqrt{\omega_\beta(t')}
    \, a_{bc}(t') e^{-i(\varphi_{t'}-\omega_b t')}
     \right|^2\,,
    \end{align}
    for plots see the supplement \cite{supplement}. For longer acceleration times the jitter decreases as $\gamma^{-3/2}$ due to the decrease of the betatron frequency via the prefactor $\omega_\beta^3(t)$. With a saddle point approximation of \eqref{eq:jitter} we find the scaling $\Delta\theta \sim L^{-3/2} \hat F_\star \gamma_p^{3/2}$, where $L$ is the acceleration length and $\hat F_\star$ is the pulse front steepness parameter at resonance. The predicted density dependence $\Delta \theta \sim n^{-3/4}$ is in qualitative agreement with the experimental findings.

    \begin{figure}[!b]
    \centering
    \includegraphics[width=\columnwidth]{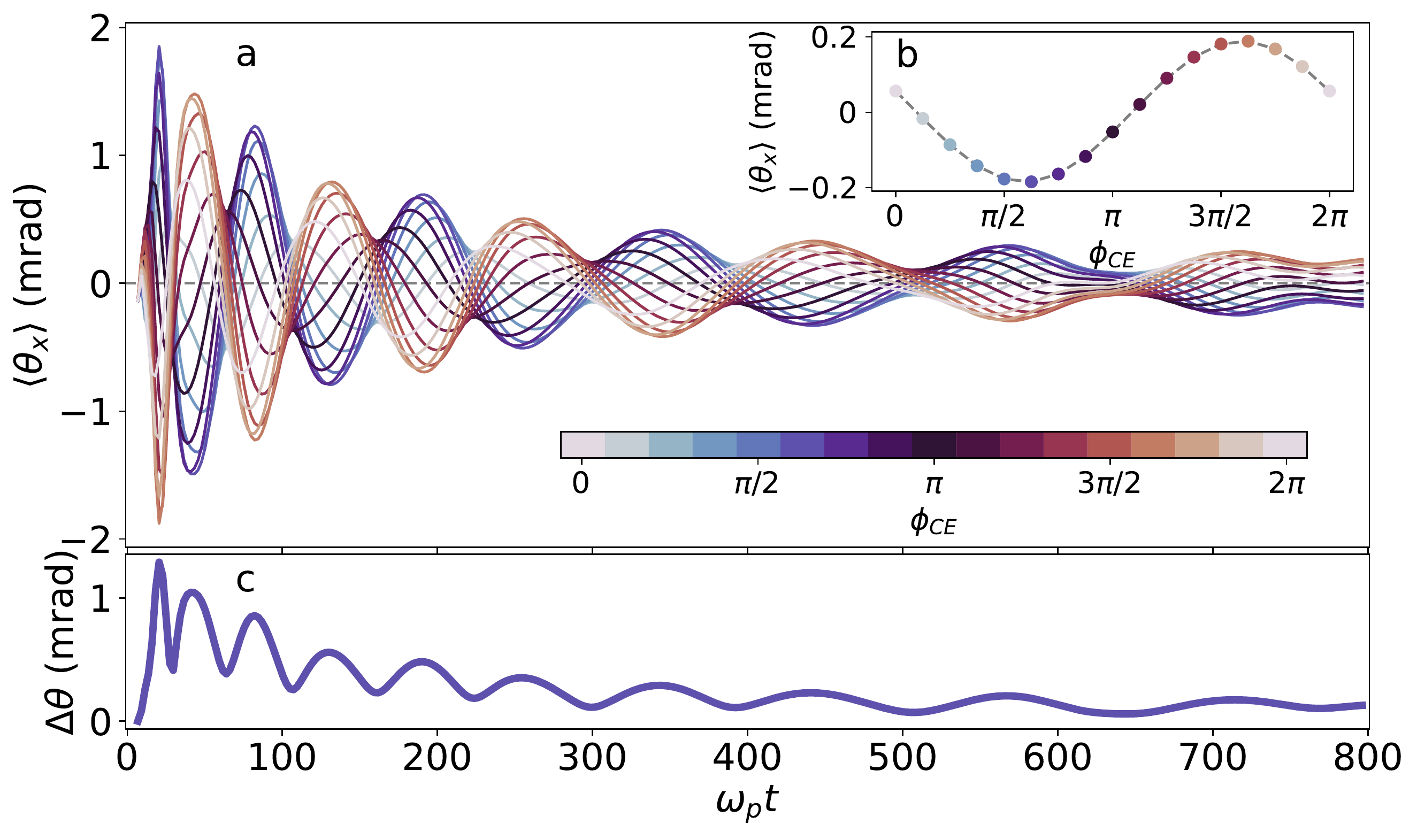}
    \caption{PIC simulation results for the CEP-dependence of the beam pointing during propagation in plasma for an externally injected beam (a), pointing at the end of the simulation at $\omega_pt=800$ (b), and evolution of rms beam jitter (c).}
    \label{fig:CEP_dependence_PICs}
    \end{figure}

    Using the betatron-wobble-resonance model, we generated a synthetic jitter diagnostic, shown in Fig.~\ref{fig:synthetic}. The parameters were chosen to give an initial pointing distribution comparable to the experimental results with the bubble wobble mechanism switched off. It shows that for reasonable parameters the polarization dependent jitter can be understood qualitatively and even quantitatively  in terms of the driven betatron oscillation. The CEP dependence of electron beam pointing is confirmed using a large set of 3D PIC simulations as shown in Fig.~\ref{fig:CEP_dependence_PICs}. In these simulations we externally injected an electron beam centrally into a wakefield to eliminate potential CEP effects from the injection. The laser pulse was defined here with a `triangular' envelope to mimic the shape of the pulse front after self-steepening. This minimizes  computational effort by eliminating the need for the pulse to self-steepen for each simulation.  Simulations with ionization injection instead of external injection and Gaussian pulses with an initial length of $c\tau= \unit{8}{\micro\metre}$ showed very similar results to those shown in Fig.~\ref{fig:CEP_dependence_PICs} (see supplement \cite{supplement}).

    Our model also allows different experimental cases a)-f) to be explained. We note that the model reproduces the general trend to smaller absolute pointing jitter for higher densities (corresponding to a longer propagation in units of $\omega_p t$ in Fig.~\ref{fig:CEP_dependence_PICs}. Also, finding a data point with very little effect (case f) and an increase again at high density (case c compared to b) is explained by  the oscillations in the magnitude of the pointing jitter with propagation length $\omega_p t$. 

    \BCO{}s can therefore contribute to both beam emittance and pointing. The emittance contribution vanishes for short injection length, but the pointing does not. Modern gas targets often are designed for localized injection~\cite{Kirchen2021_HamburgCap}. However, this is insufficient to achieve  stable pointing as shown in Figure \ref{fig:CEP_dependence_PICs}. However for constant initial CEP all phases are constant and the injection direction is also reproducible shot to shot. CEP control is therefore necessary to suppress the porlarization dependent pointing. We note that longer drive pulses do not, in general, eliminate the CEP dependent effect, as a steep pulse-fronts occur even for initially long pulses, due to the nonlinear pulse evolution in the plasma. However, controlling the CEP phase to within \unit{500}{\milli\rad} \cite{Golinelli:19} as shown in Fig. \ref{fig:CEP_dependence_PICs} b) constrains the  polarization induced jitter to  below \unit{50}{\micro\rad}.

    In addition to the excitation of collective betatron oscillations discussed here, there could be additional CEP dependent contributions if the driving laser pulse contains single-cycle features already during injection of the electrons, as discussed recently for the case of near-single-cycle pulses \cite{Huijts:PhysPlasmas2021,Huijts:2022waveform}.

    In conclusion we have experimentally and theoretically identified a mechanism intrinsic to the LWFA process that couples both CEP and porlarization to the electron beam pointing. This mechanism is fundamental to laser wakefield accelerators in the sense that the regime of pulse front etching and steepening is necessary for the efficient operation of LWFAs and no means of eliminating the instability is obvious at this time. Stable operation can nonetheless be achieved if the phase of the oscillation at the output of the accelerator is kept constant, requiring a CEP-stable drive laser, and a sufficiently reproducible target density profile and injection region.


The authors gratefully acknowledge the Gauss Centre for Supercomputing e.V. (www.gauss-centre.eu) for funding this project by providing computing time through the John von Neumann Institute for Computing (NIC) on the GCS Supercomputer JUWELS at Jülich Supercomputing Centre (JSC). Particle-in-cell simulations were performed with SMILEI~\cite{smilei}.

\nocite{Ma:PhysPlas2018,Ferri:PRAB2018,emcee,kim2017stable}
\bibliography{references}

\end{document}


\title{Polarization dependent beam pointing jitter in laser wake field accelerators \protect\\ Supplemental Material}
\input{authors.tex}

\maketitle

\listoffigures

\section{Laser Pointing Fluctuations}
    
As can be seen in Figure~1 of the manuscript there is a trend for the electron beam  pointing jitter to be greater  horizontally than vertically. This effect is due to the laser pointing fluctuating of \jeti being stronger by a factor of 6 in the horizontal plane at the laser output. These pointing fluctuations are much smaller than the electron beam pointing and on the order of 10s of $\micro\rad$. Laser pointing fluctuations are known to be  magnified by transverse refractive index gradients in plasma density gradients at the entry and exit of the gas cell, which act as defocusing lenses \cite{Ma:PhysPlas2018,Ferri:PRAB2018}.

\section{Electron beam profile and spectrum}
Figure \ref{fig:Beamspot}a) shows a typical beam profile of an electron bunch from the data set for figure~1. A low ellipticity and a larger halo can be seen, which are due to the laser and density parameters in the gas jet selected for minimizing the electron pointing jitter. However, choosing a different set of parameters can lead to a smaller and less elliptic beam profile as can be seen in figure \ref{fig:Beamspot}b). The aforementioned parameters for optimization include the gas density, the position of the focal spot relative to the gas target, and the GDD of the laser with the latter also having a substantial influence on the electron energy spectrum \cite{kim2017stable}.

\section{Bayesian Inference}

In addition to the ratio-of-ellipses measure $\mathcal R$ discussed in the main text, Bayesian inference was used to estimate the value of the polarization dependent jitter contribution~\cite{emcee}. The generative model for the data is Gaussian with tilted covariance ellipses, in which the polarization dependent jitter contribution $\sigma_P$ is attributed to either $H$ or $V$ depending on the laser polarization direction being horizontal or vertical. The unpolarized jitter contribution is assumed to have the same magnitude along the major and minor axes for both $H$ and $V$ cases, but we allow for a relative tilt of those axes for $H$ and $V$ independently. The data are the tuples $D = (x_H,y_H,x_V,y_V)$, where over each dataset the average position is normalized to zero, i.e they represent the residuals $x_H=X_H- \mu_H$, etc.

The log likelihood is given by
\begin{align}
\ln \mathcal L(D|\Theta) 
=
- \frac{1}{2}\sum_{\rm data}\left[
        (x_{H},y_{H}) C_H^{-1} (x_{H},y_{H})^T
    +   (x_{V},y_{V}) C_V^{-1} (x_{V},y_{V})^T
    + \ln \det (C_H)
    + \ln \det (C_V)
    \right]
\end{align}
with the horizontal and vertical covariance matrices
\begin{align}
C_{H} & =
\frac{1}{2}
\left(
    \begin{matrix}
    \sigma_a^2 + \sigma_b^2 + (\sigma_a^2-\sigma_b^2)\cos 2\varphi_H + 2\sigma_P^2& 
    (\sigma_a^2-\sigma_b^2)\sin2\varphi_H \\
    (\sigma_a^2-\sigma_b^2)\sin2\varphi_H &
    \sigma_a^2 + \sigma_b^2 - (\sigma_a^2-\sigma_b^2)\cos 2\varphi_H
  \end{matrix}
\right)\,,\\
C_{V} &=
\frac{1}{2}
\left(
    \begin{matrix}
    \sigma_a^2 + \sigma_b^2 + (\sigma_a^2-\sigma_b^2)\cos 2\varphi_V & 
    (\sigma_a^2-\sigma_b^2)\sin2\varphi_V \\
    (\sigma_a^2-\sigma_b^2)\sin2\varphi_V &
    \sigma_a^2 + \sigma_b^2 - (\sigma_a^2-\sigma_b^2)\cos 2\varphi_V + 2\sigma_P^2
  \end{matrix}
\right)\,.
\end{align}

\begin{figure}[!ht]
    \centering
    \includegraphics[width=\textwidth]{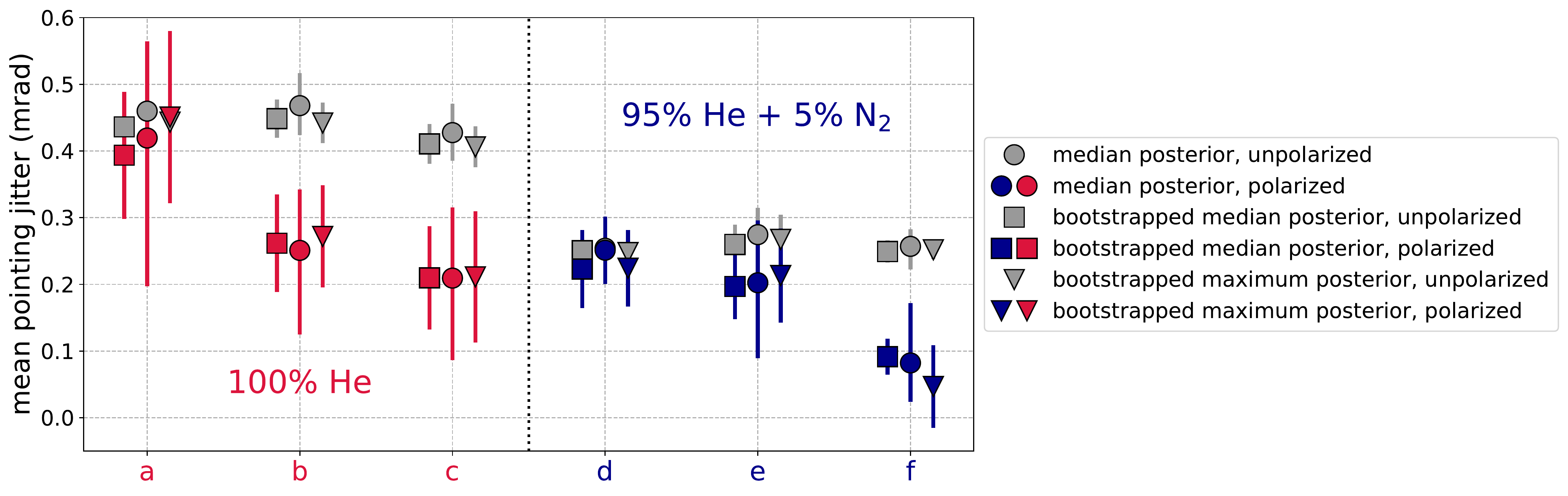}
    \caption[Bayesian inference results]{
    Markov-chain Monte Carlo Bayesian inference of the the polarization dependent contribution to the pointing jitter for the datasets (a-f) [cf. Figure~1 of the manuscript] as red/blue symbols; Grey symbols are the polarization \emph{independent} jitter contribution as the geometric mean of the major and minor axes $\sqrt{\sigma_a\sigma_b}$.
    Circles are the median of the posterior distribution (errorbars range from 15th to 85th percentile). Square and triangle symbols are for a bootstrapping analysis where squares are the bootstrapped median and triangles the bootstrapped maximum posterior. Errorbars represent the standard deviation of the mean the latter two cases.
    }
    \label{fig:ExperimentalData_Markov}
\end{figure}

Results from the Bayesian inferrence are summarized in Figure~\ref{fig:ExperimentalData_Markov}. The polarization independent jitter contribution, defined as the geometric mean $\sqrt{\sigma_a\sigma_b}$ has a constant value of approximately $\unit{0.44}{\milli\rad}$ for 100 \% He gas, while for the gas mixture with ionization injection it is smaller at approximately $\unit{0.26}{\milli\rad}$. We found the polarization dependent contribution to the beam jitter $\sigma_P$ in some datasets (a,d,e) being almost as large as the unpolarized jitter contribution. With increasing plasma density the polarization dependent jitter contribution decreases, while the polarization independent jitter remains constant. For dataset (f) the Bayesian analysis shows no significant polarization dependent jitter, which is consistent with the analysis in Figure~1 of the manuscript.


\begin{figure}[p]
    \centering
    \includegraphics[width=0.9\columnwidth]{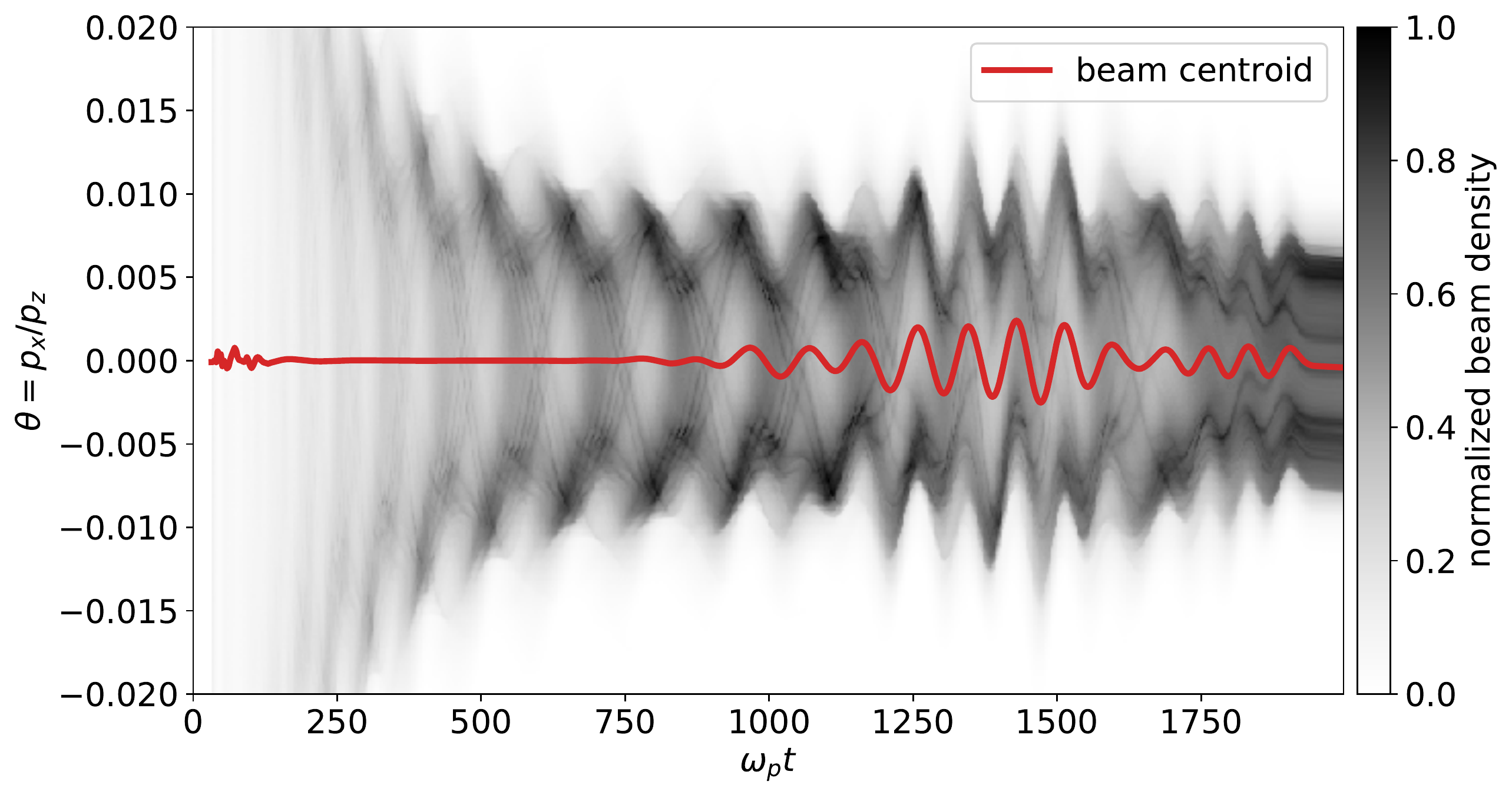}
    \caption[PIC results for beam centroid oscillation for a Gaussian pulse]{Results of a PIC simulation with a temporally Gaussian pulse with initial duration of $c\tau=\unit{8}{\micro\metre}$ and $a_0=4$ showing that the beam centroid starts oscillating eventually also for initially Gaussian pulses. The plot exhibits the density of the externally injected beam as a function of the pointing angle. After an initial phase where the beam centroid pointing (red curve) is stable at $\theta=0$, the pointing jitter grows rapidly once the front edge has steepened, exciting bubble centroid oscillations resulting in large the beam centroid oscillation due to forced collective betatron oscillations when resonance is reached.}
    \label{fig:PIC_Gaussian}
\end{figure}

\begin{table}[p]
    \centering
    \caption{PIC simulation configuration: The simulation parameters for the runs with a triangular pulse shape shown in the manuscript are summarized in this table. Figure \ref{fig:PIC_Gaussian} of this supplement show the results of a simulation for \jeti parameters with an initially Gaussian pulse shape, demonstrating that the instability occurs for typical LWFA pulse parameters and longer initial pulses in agreement with the experimental results.}
    \label{tab:PIC_config}
    \begin{tabular}{p{4cm}|p{4cm}}
    \toprule
         \multicolumn{2}{c}{\textbf{Laser parameters}}  \\ \midrule
         Wavelength, $\lambda_L$     & $\unit{0.8}{\micro\metre}$ \\  
         Spotsize, $w_0$             & $\unit{14}{\micro\metre}$ or $4/k_p$ \\ 
         Strength, $a_0$             & $4$    \\ 
         Linear rising edge                 & $\unit{2}{\micro\metre}$ (FWHM) \\  
         Linear falling edge                 & $\unit{8}{\micro\metre}$ (FWHM) \\ 
         CEP phase                   & $0 \leq \phi_{CE} \leq 2 \pi$ \\ \midrule
         \multicolumn{2}{c}{\textbf{Plasma parameters}}  \\ \midrule
         Matched plasma density & $\unit{2.4\times 10^{18}}{\centi\metre^{-3}}$ or $\gamma_p=27.48$ \\ 
         Density profile             & $\unit{50}{\micro\metre}$ upramp, then uniform \\ \midrule
         \multicolumn{2}{c}{\textbf{Beam parameters}} \\ \midrule
         %
         Initial beam size &  $\unit{1}{\micro\metre}$ \\ 
         Spatial profile & Gaussian \\ 
         Initial $gamma$   & (0, $\gamma_p/2$)=(0.0, 13.7) \\ 
         Initial trans. offset &  $x_0 =0.0$  \\ 
         Initial trans. angle  & $\theta_{x0} =0.0$ \\ 
         Initial temperature       &  0.08  \\ \midrule
         \multicolumn{2}{c}{\textbf{Numerical configuration}} \\ \midrule
         Simulation window, $L_x \times L_z$   &  $4.6/k_p \times 2.3/k_p$ \\ 
         Grid resolution, $\Delta x \times \Delta z$ &  $32/\lambda_L \times 32/\lambda_L$ \\ \bottomrule
     \end{tabular}
\end{table}

    \begin{figure}[!ht]
    \centering
    \includegraphics[width=0.6\columnwidth]{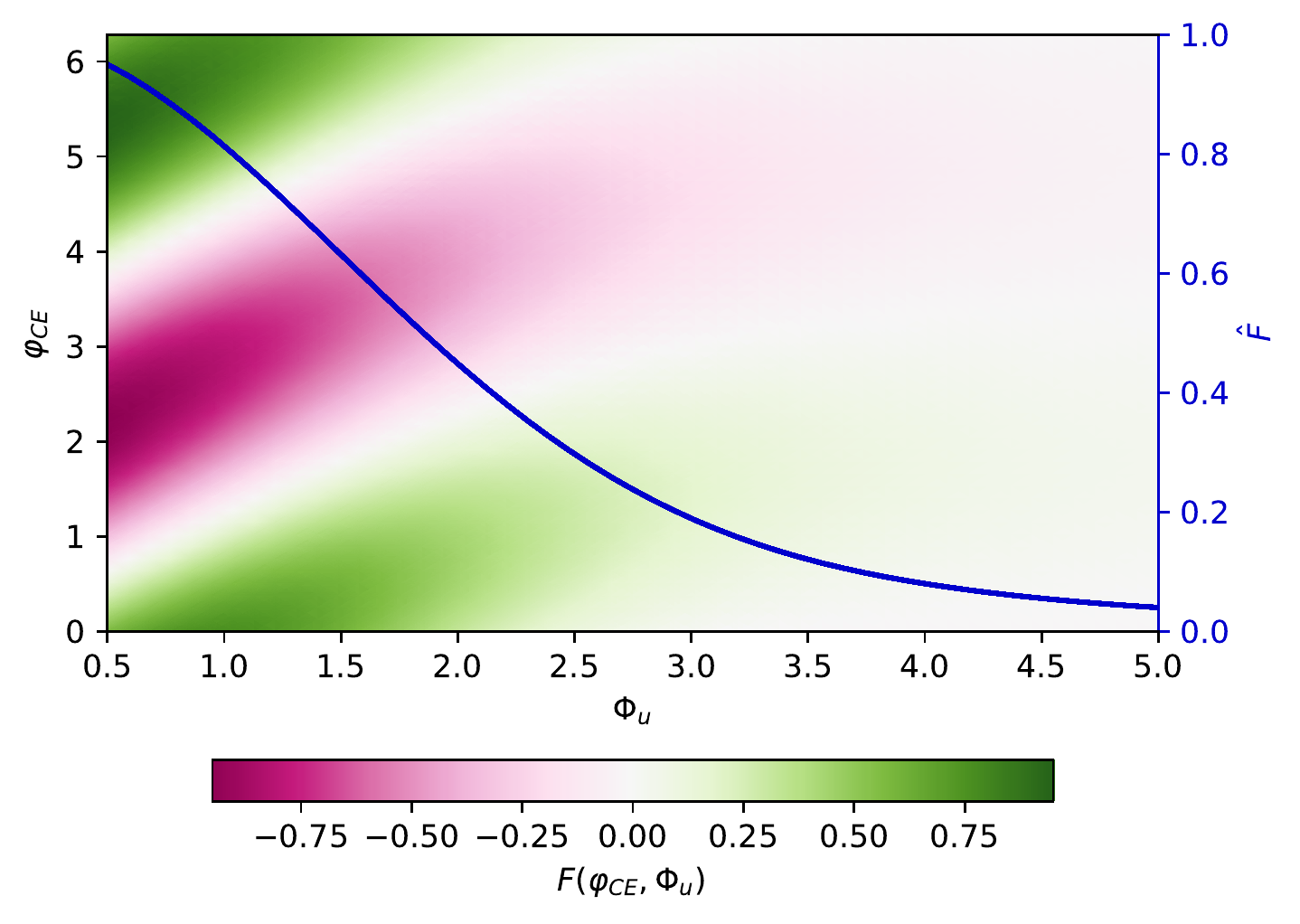}
    \caption[Plot of the pulse front steepness function $F$]{Contour plot of the functions $F$ and $\hat F =\max_{\phi_{CE}} F$ determining the amplitude of bubble centroid oscillations. Here $f=h \cos(\zeta+\phi_{CE})$ with the envelope $h(\zeta) = e^{-\zeta^2/2\Phi_u} \theta(\zeta) + e^{-\zeta^2/2\Phi_d}\theta(-\zeta)$, and step function $\theta(.)$. We use a short pulse front duration $\Phi_u$ and a long pulse back duration $\Phi_d=25$. Note that the result is nearly independent of $\Phi_d$ for $\Phi_d > 5$.
    }
    \label{fig:upramp_asy}
    \end{figure}
    
    \begin{figure}[p]
    \centering
    \includegraphics[width=0.8\textwidth]{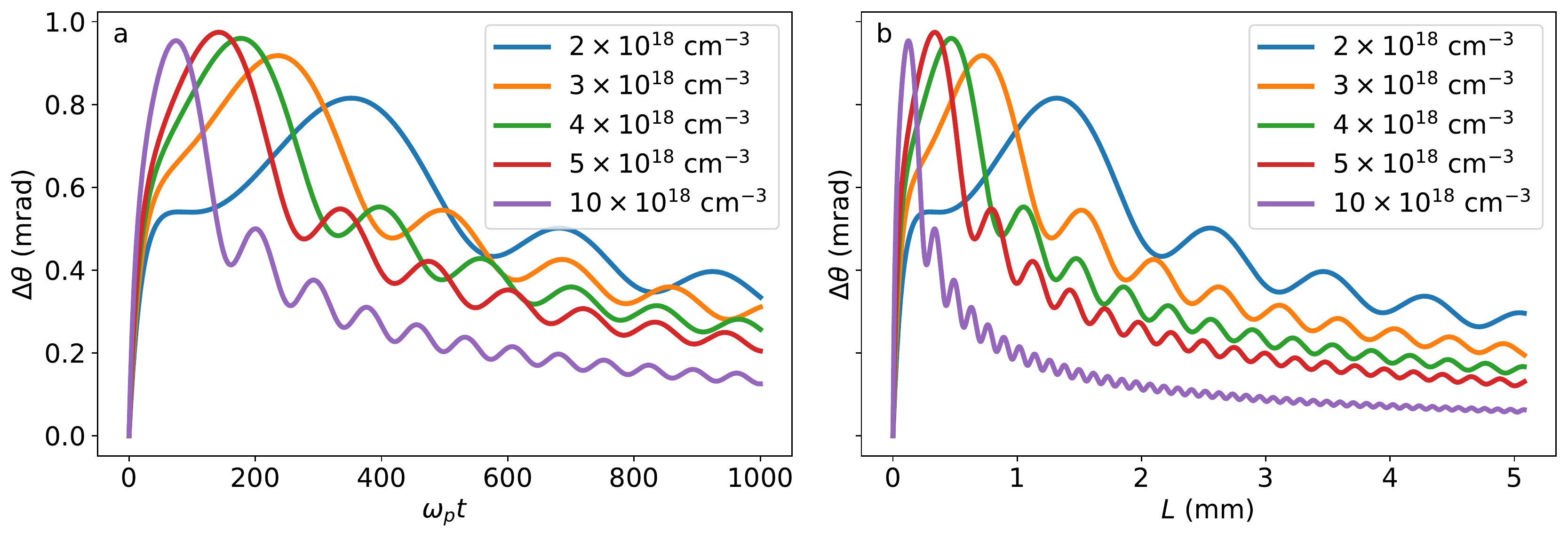}
    \includegraphics[width=0.8\textwidth]{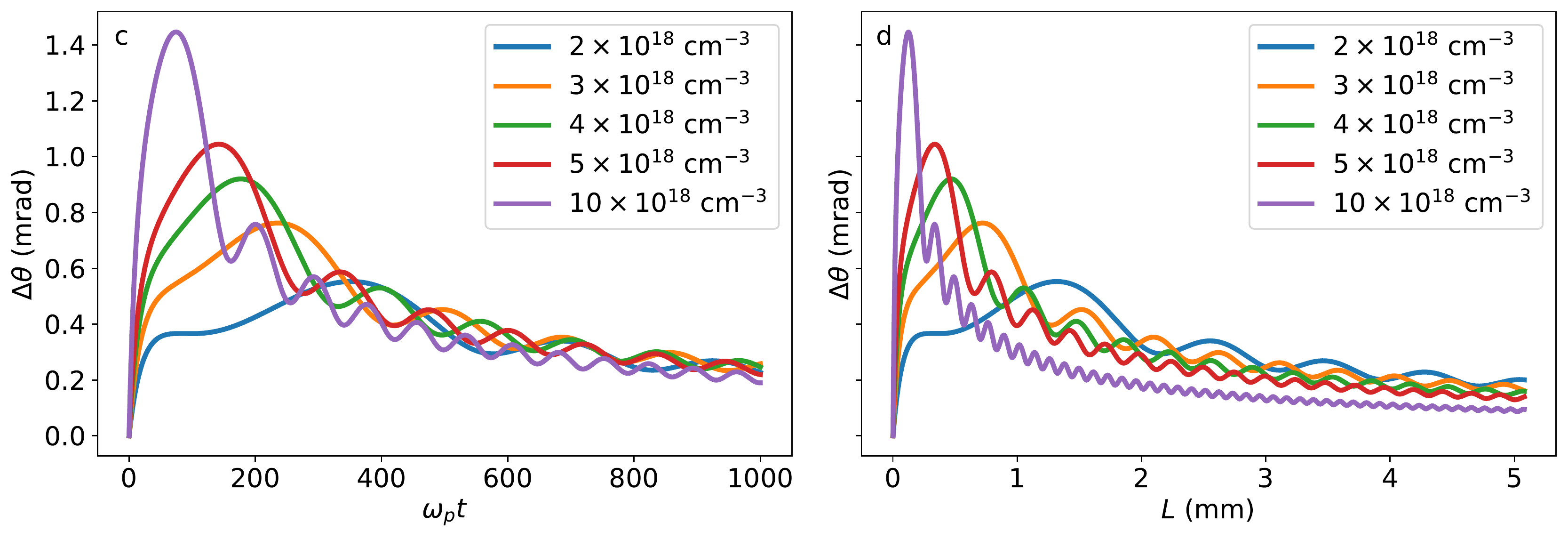}
    \caption[Plots of the analytic result for the jitter in Eq.~(5)]{Plot of the analytic result of the pointing jitter as given by Eq.~(5) in the manuscript. Results are shown for fixed $a_{bc}=0.005$ at various plasma densities (top panels) and for the bubble centroid oscillation amplitude scaling with density $a_{bc}=0.1/\gamma_p$ (lower panels). For convenice we show the results both as a function of normalized $\omega_p t$ (left panels), as well as a function of the acceleration length in millimetres (right panels). All results show fast initial increase of pointing jitter due to the betatron-BCO-resonance before the $\Delta \theta$ slowly decreases as the electrons are further accelerated and the betatron frequency drops as $\omega_\beta(t) \propto 1/\sqrt{2\gamma(t)}$.}
    \label{fig:analytic_jitter}
    \end{figure}

\begin{figure}[!htp]
    \includegraphics[width=0.45\textwidth]{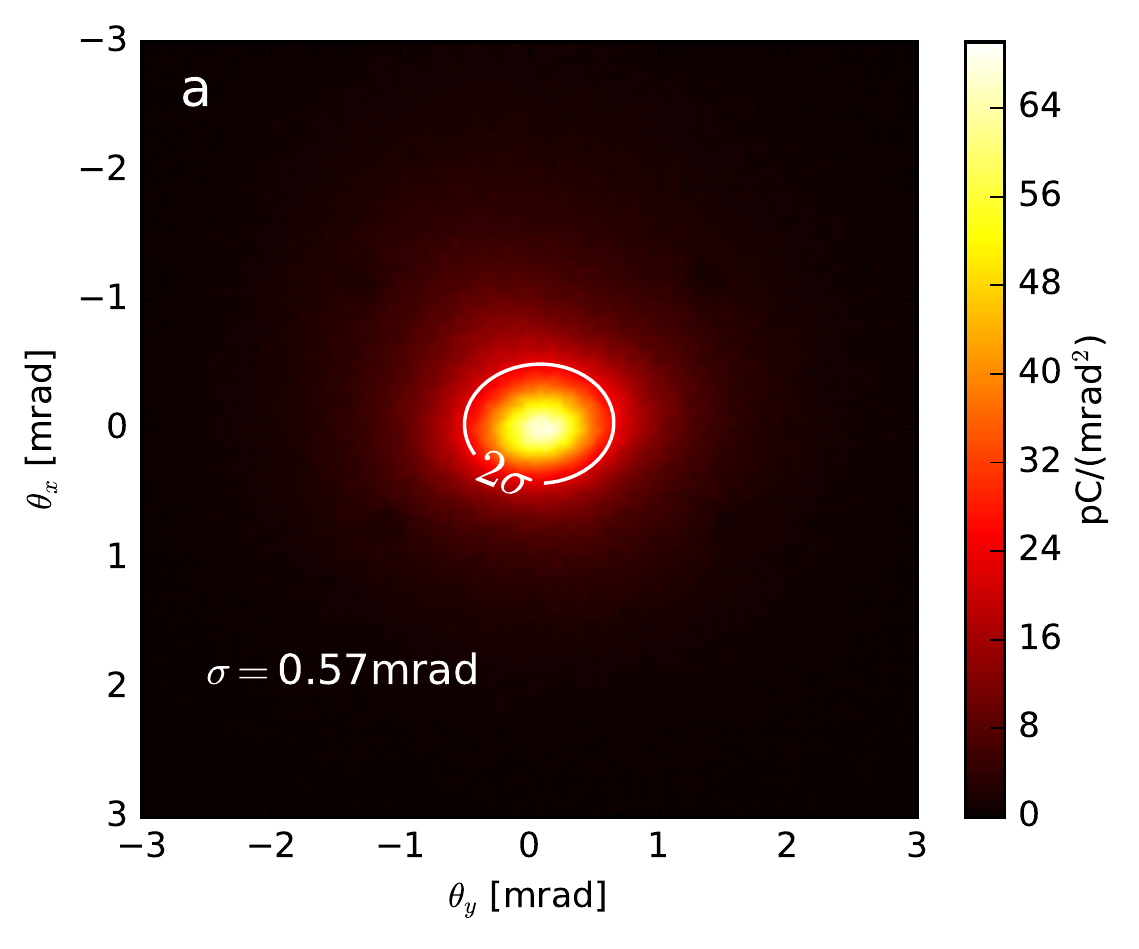}
    \includegraphics[width=0.45\textwidth]{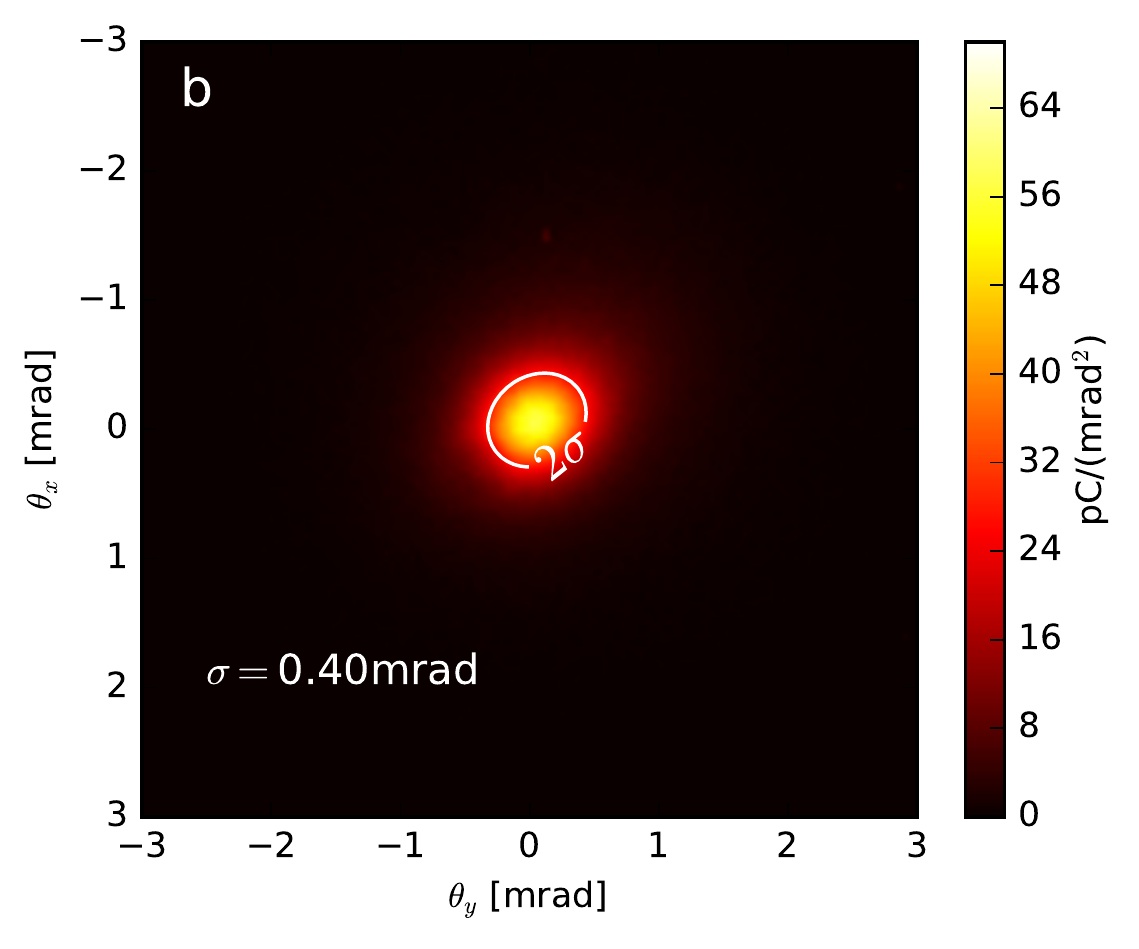}
    \caption[Experimental electron beam profiles]{Representative electron beam profile with $\sigma$ indicating the electron spot radius. (a) Profile for data set in Figure 1 and (b) with optimized gas density profile for low divergence
    }
    \label{fig:Beamspot}
\end{figure}

\begin{figure}[h]
    \centering
    \includegraphics[width=0.4\columnwidth]{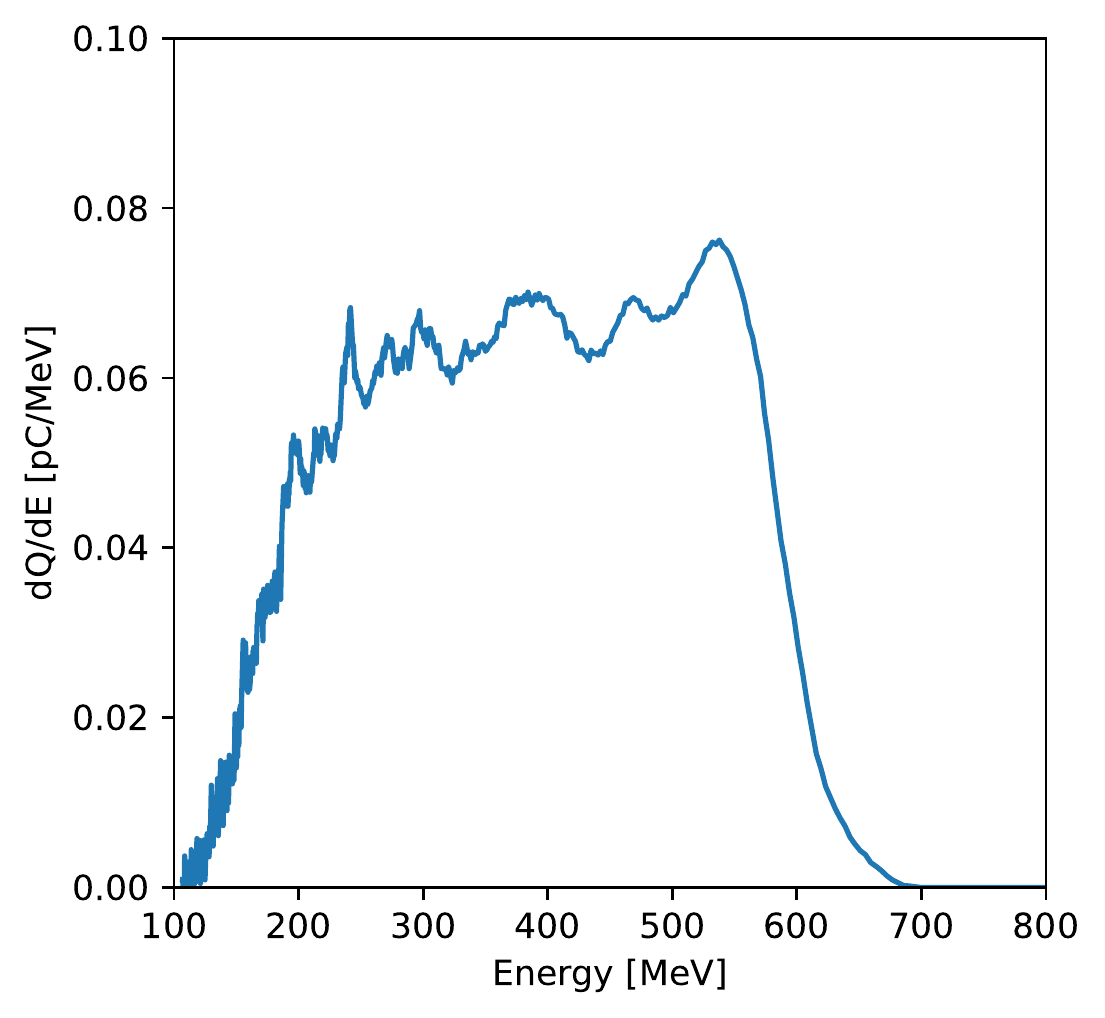}
    \vspace{-0.5cm}
    \caption[Electron spectrum for self-injection]{Representative electron beam spectrum for self-injection.
    }
    \label{fig:Spec_Selfinjection}
\end{figure}
\vspace{0.5cm}
\begin{figure}[b]
    \centering
    \includegraphics[width=0.4\columnwidth]{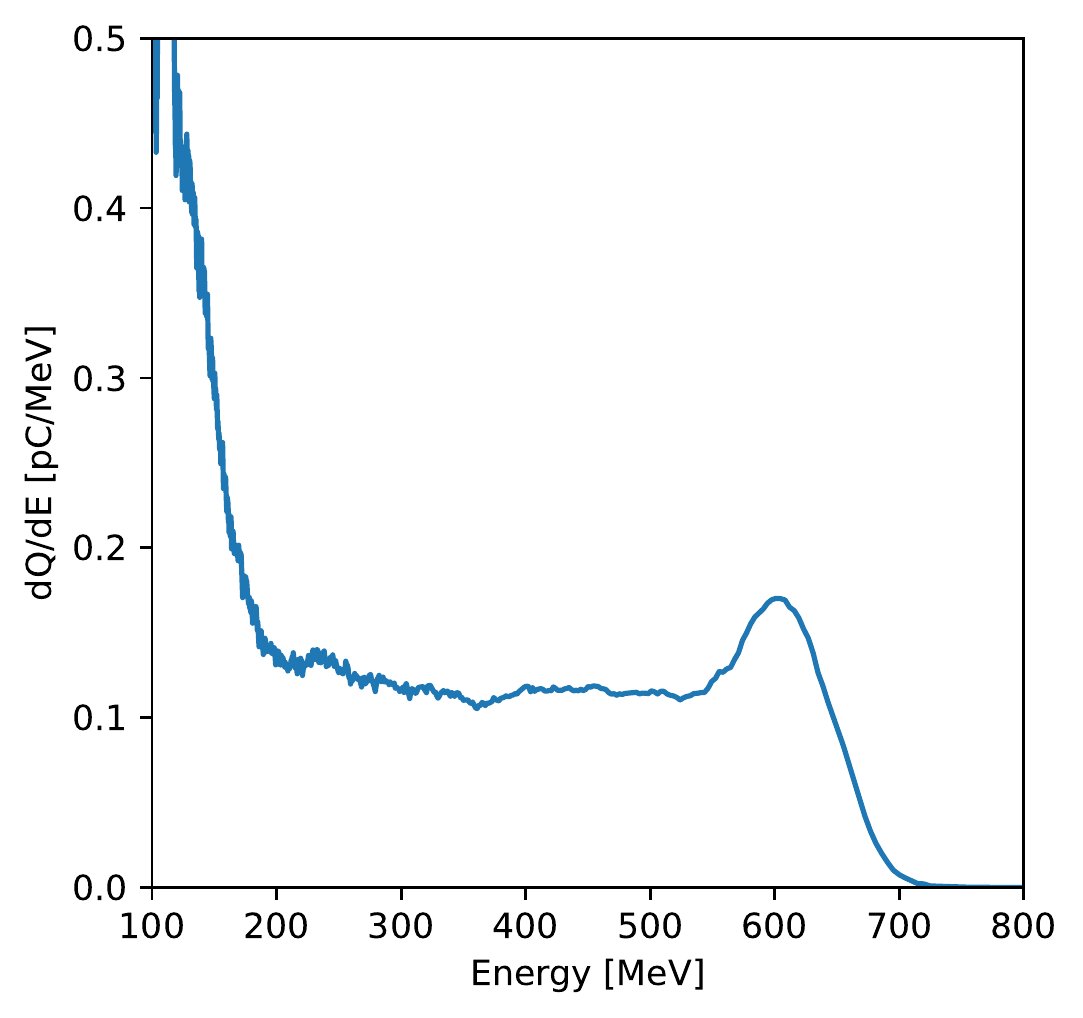}
    \vspace{-0.5cm}
    \caption[Electron spectrum for ionization injection]{Representative electron beam spectrum for ionization-injection.
    }
    \label{fig:BSpec_Ionizationinjection}
\end{figure}

\bibliography{supplement}

%% file: authors.tex
\newcommand{\IOQ}{Institute of Optics and Quantum Electronics, Max-Wien-Platz 1, 07743 Jena, Germany}
\newcommand{\HIJ}{Helmholtz-Institute Jena, Fröbelstieg 3, 07743 Jena, Germany}
\newcommand{\GSI}{GSI Helmholtzzentrum für Schwerionenforschung GmbH, Planckstraße 1, 64291 Darmstadt, Germany}

\author{A. Seidel}
\email{seidel.andreas@uni-jena.de}
\affiliation{\IOQ}
\affiliation{\HIJ}

\author{B. Lei}
\email{b.lei@hi-jena.gsi.de}
\affiliation{\IOQ}
\affiliation{\HIJ}
\affiliation{\GSI}

\author{C. Zepter}
\affiliation{\IOQ}
\affiliation{\HIJ}

\author{M. C. Kaluza}
\affiliation{\IOQ}
\affiliation{\HIJ}
\affiliation{\GSI}

\author{A. Sävert}
\affiliation{\IOQ}
\affiliation{\HIJ}
\affiliation{\GSI}

\author{M. Zepf}
\affiliation{\IOQ}
\affiliation{\HIJ}
\affiliation{\GSI}

\author{D. Seipt}
\affiliation{\HIJ}
\affiliation{\GSI}